%% file: planglow.tex
\documentclass[sigconf]{acmart}

\AtBeginDocument{%
  }

\copyrightyear{2025} 
\acmYear{2025} 
\setcopyright{cc}
\setcctype{by}
\acmConference[L@S '25]{Proceedings of the Twelfth ACM Conference on Learning @ Scale}{July 21--23, 2025}{Palermo, Italy}
\acmBooktitle{Proceedings of the Twelfth ACM Conference on Learning @ Scale (L@S '25), July 21--23, 2025, Palermo, Italy} \acmDOI{10.1145/3698205.3729541}
\acmISBN{979-8-4007-1291-3/2025/07}

\usepackage{graphicx}
\usepackage{xcolor}
\usepackage{fontawesome5}
\usepackage{multirow} 
\usepackage{color}
\usepackage{soul}
\usepackage{balance} 

\begin{document}

\title{PlanGlow: Personalized Study Planning with an Explainable and Controllable LLM-Driven System}

\author{Jiwon Chun}
\affiliation{%
  \institution{Texas A\&M University}
  \city{College Station}
  \state{Texas}
  \country{USA}}
\email{jiwonchun@tamu.edu}

\author{Yankun Zhao}
\affiliation{%
  \institution{WisdomPlan Inc.}
  \city{Pittsburgh}
  \state{Pennsylvania}
  \country{USA}
}
\email{yankunzhao@wisdomplan.ai}

\author{Hanlin Chen}
\affiliation{%
 \institution{Vanderbilt University}
 \city{Nashville}
 \state{Tennessee}
 \country{USA}}
\email{hanlin.chen@vanderbilt.edu}

\author{Meng Xia}
\affiliation{%
  \institution{Texas A\&M University}
  \city{College Station}
  \state{Texas}
  \country{USA}}
\email{mengxia@tamu.edu}

\renewcommand{\shortauthors}{Chun et al.}

\begin{abstract}
  Personal development through self-directed learning is essential in today's fast-changing world, but many learners struggle to manage it effectively. While AI tools like large language models (LLMs) have the potential for personalized learning planning, they face issues such as transparency and hallucinated information. To address this, we propose \textit{PlanGlow}, an LLM-based system that generates personalized, well-structured study plans with clear explanations and controllability through user-centered interactions. Through mixed methods, we surveyed 28 participants and interviewed 10 before development, followed by a within-subject experiment with 24 participants to evaluate \textit{PlanGlow}'s performance, usability, controllability, and explainability against two baseline systems: a GPT-4o-based system and Khan Academy’s Khanmigo. Results demonstrate that \textit{PlanGlow} significantly improves usability, explainability, and controllability. Additionally, two educational experts assessed and confirmed the quality of the generated study plans. These findings highlight \textit{PlanGlow}'s potential to enhance personalized learning and address key challenges in self-directed learning.
\end{abstract}



\begin{teaserfigure}
    \centering
    \includegraphics[width=1\linewidth]{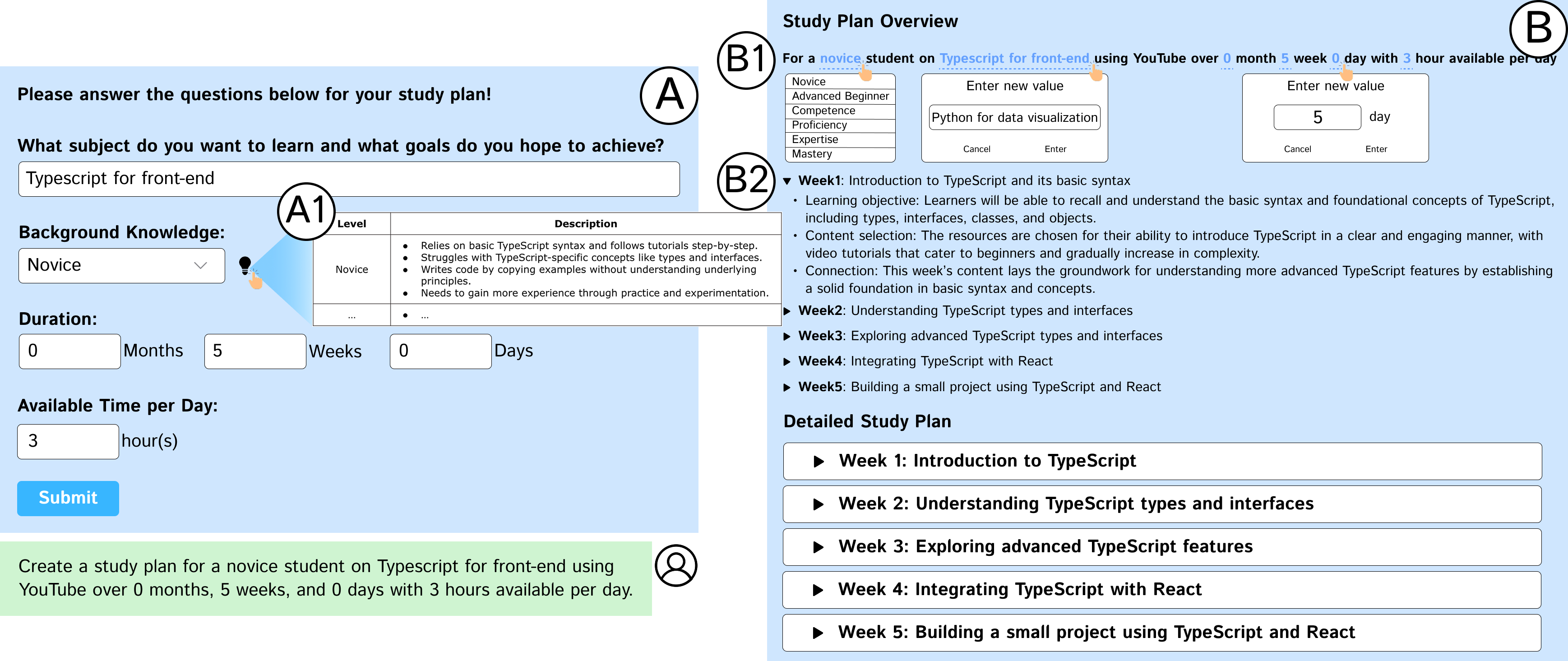}
    \caption{\textit{PlanGlow} contains two major components. (A) is the plan generation interface where users specify their learning preferences such as subject, prior knowledge, and available time. (A1) is the background level description. (B) is the generated plan interface that presents a personalized study plan, organized into weekly segments with a detailed daily breakdown. (B1) allows in-line editing of learning goals, background knowledge level, duration, and daily availability. (B2) is the weekly breakdown including learning objectives, the reasons behind content selection, and key connections within the weekly.}
    \label{fig:teaserl}
\end{teaserfigure}


\maketitle

\input{sections/1-intro}
\input{sections/2-related}
\input{sections/3-user_needs}
\input{sections/4-system}
\input{sections/5-evaluation}
\input{sections/6-results}

\input{sections/7-discussion}
\input{sections/8-conclusion}

\bibliographystyle{ACM-Reference-Format}
\balance
\bibliography{planglow}
\end{document}

%% file: sections/1-intro.tex
\section{Introduction}
Self-directed learning is a process where individuals take the initiative in their learning, with or without external assistance~\cite{boyer2014self,lin2024exploring}. Effective self-directed learning can enhance their overall learning experience, cultivate their habits of continuous learning, and promote personal growth~\cite{song2016motivational, lu2024exploring,park2024promise, li2023analysis,kim2014leveraging,lin2024exploring}. Learning platforms such as MOOCs, including edX\footnote{\href{https://www.edx.org}{https://www.edx.org}}, aim to support self-directed learners at scale with massive learning resources, but often lack individualized guidance.

It is challenging to provide personalized support at various stages of the self-directed learning process, from goal setting, planning, and selecting appropriate resources~\cite{li2023analysis} to finding and effectively using interactive and multimedia resources~\cite{lin2024exploring}. Emerging technologies address these challenges by enabling personalized learning with access to information anytime, anywhere, making the process more accessible, autonomous, active, and independent~\cite{rafiola2020effect, chictiba2012lifelong, zhang2020understanding}. Among these technologies, large language models (LLMs) have advanced self-directed learning in different stages~\cite{fiok2022explainable,limo2023personalized,lin2024exploring,li2024reconceptualizing}. 

Most research applies LLMs during the execution stage of self-directed learning, providing resource access and task support. For instance, Ali et al. developed a GPT-4-based chatbot for outside-class Q\&A and self-assessment through one-to-one interactions~\cite{ali2023supporting}. However, research on the planning stage, which involves goal setting, plan generation, and resource location, remains limited~\cite{li2023analysis}. Although Lin et al. examined ChatGPT's role in supporting adult learners during the planning process~\cite{lin2024exploring}, comprehensive solutions that address these challenges are still lacking.

In this context, LLM-based tools for the planning stage of self-directed learning face two major challenges that can impact user experience and effectiveness~\cite{memarian2023chatgpt}. The first is the lack of transparency, which makes it difficult for learners to trust and follow recommendations without clear explanations~\cite{memarian2023chatgpt, schoenherr2023designing}. The second is the potential for incorrect information, which can confuse learners, negatively affecting their overall experience. The system can produce artificial hallucination information that lacks real-world basis~\cite{alkaissi2023artificial}. These inaccuracies can further disrupt the planning process and require additional effort from learners to verify and correct.

To address the limitations of current LLM-based tools, we introduce \textit{PlanGlow}, an LLM-assisted system for self-directed learning planning. Through a literature review, we identified explainability and controllability as key factors. Explainability provides clear justifications for AI recommendations, fostering trust and informed interaction~\cite{schoenherr2023designing,zhao2024explainability,zhang2023language}, while controllability allows users to adjust outputs and correct errors like hallucinations~\cite{ji2023survey}. These two factors are interconnected because explainability provides users with the reasoning behind AI output, enhancing their ability to refine system behavior and reduce errors or hallucinations~\cite{zhang2023language,ji2023survey}. With these features, \textit{PlanGlow} contributes to scalable educational tools that support effective self-directed learning planning.

To validate these features, we conducted surveys and interviews with users after they generated study plans with GPT-4o. Based on this feedback, we consulted an educational expert to establish our theoretical framework such as Bloom's Taxonomy for learning objectives~\cite{uark_blooms_taxonomy} and cognitive load theory~\cite{brown1978knowing, duell1986mcskills, flavell1976metacognitive} for content organization. These insights guided the development of \textit{PlanGlow}, which integrates explainability and controllability as core features. Explainability offers concise summaries of the content, learning objectives, rationale for the chosen content, and connections between weekly and daily plans. Controllability lets users iteratively adjust preferences, validate resources, and refine study materials, aligning with Universal Design principles for diverse needs~\cite{burgstahler2009universal}. 

This dual approach improves user control, trust, and reliability by reducing misinformation and hallucinations. We conducted user studies to compare \textit{PlanGlow} with two baselines: a GPT-4o-based system and Khan Academy's educational coach, Khanmigo. The results indicated significant improvements in both controllability and explainability, highlighting \textit{PlanGlow}'s effectiveness.

The contributions of this work are summarized as follows. 
\begin{itemize}
\item We propose \textit{PlanGlow}, an LLM-based personalized study planning system designed to support learners in the initial stages of self-directed learning. By integrating explainability and controllability, \textit{PlanGlow} provides a transparent and adaptive educational experience, encouraging learners to understand recommendations, make informed decisions, and take ownership of their learning path. The source code is available on Github\footnote{\href{https://github.com/dreamlab-24/PlanGlow}{https://github.com/dreamlab-24/PlanGlow}}.
\item We conduct a within-subject experiment to evaluate the performance, usability, explainability, and controllability of \textit{PlanGlow}, in comparison with two baseline
systems. Experiment results show that users find \textit{PlanGlow} more intuitive and integrated than the baselines, with controllability to create plans, verify resources, and explore alternatives. It also provides clear explanations, aligns plans with user goals, and clarifies task connections, allowing for more informed decision making.
\end{itemize}

%% file: sections/2-related.tex
\section{Related Work} 
This section reviews LLM-based self-directed learning and explores explainability and controllability in educational AI systems.

\subsection{Use of LLMs in Self-Directed Learning}
Recent studies have explored using LLMs for self-directed learning such as progress monitoring, and reflection~\cite{lemmetty2020self, lin2024exploring,cain2024prompting,ali2023supporting,jones2022capturing}.
However, most studies focus on execution and monitoring phases, emphasizing real-time assessments, feedback, and support~\cite{askarbekuly2024llm,jones2022capturing,esiyok2024acceptance,ali2023supporting,bosch2024ai,lin2024exploring,cain2024prompting}. For instance, chatbots serve as virtual tutors, helping learners reflect on their understanding and adjust learning strategies~\cite{esiyok2024acceptance,ali2023supporting,bosch2024ai}. TeacherGAIA, a GPT-4-based chatbot, illustrates this by guiding students through constructivist processes such as knowledge construction, inquiry-based learning, self-assessment, and peer teaching~\cite{ali2023supporting}. LLMs also enhance assessment by evaluating students' performance and delivering personalized feedback~\cite{askarbekuly2024llm,jones2022capturing}. Jones et al. developed a platform that enables teachers to design tasks and guide LLMs to provide formative feedback that identifies misunderstandings and encourages reflection~\cite{jones2022capturing}.

However, a gap still remains in applying LLMs to the planning phase of self-directed learning. This phase, which includes goal setting, schedules structuring, and resource selection, is critical but underexplored. While previous works suggest that LLM-based tools such as ChatGPT could assist adult learners by identifying needs, setting goals, and recommending resources~\cite{lin2024exploring,firat2023chat}, they often remain broad and risk delivering hallucination information. Consequently, more targeted research is needed to develop structured, personalized learning plans and address LLMs' limitations.

\subsection{Explainability in AI for Education}
LLMs have limitations in self-directed learning including biases, reliance on user input, lack of transparency, and hallucinations~\cite{gallegos2024bias,jones2022capturing,he2019practical,topol2019high,lin2024exploring,cain2024prompting}. These issues highlight the need for trustworthy AI systems that are transparent, robust, and secure to ensure ethical, legal, and safety standards are upheld~\cite{markus2021role}. Explainable AI (XAI) addresses this by clarifying AI reasoning, thus fostering trust, enhancing user satisfaction, and supporting metacognitive processes~\cite{chen2017explaining,markus2021role,mittelstadt2019principles,kay2001learner}. 

However, flawed explanations can lead to biases, confusion, and disrupting learning~\cite{kendeou2023nature,fiok2022explainable}. Several XAI systems already support self-directed learning, such as FUMA, which delivers hierarchical explanations in open-ended environments, improving learning outcomes and trust~\cite{conati2021toward,khosravi2022explainable,knight2020augmenting,knight2020you,kardan2015providing}. While the importance of explainability is recognized for LLM-driven planning tools~\cite{khosravi2022explainable}, research on effectively implementing and measuring XAI during the planning phase remains limited \cite{lin2024exploring}.

\subsection{Controllability in AI for Education}
Controllable AI is an emerging strategy for managing real-world AI limitations by allowing users to guide systems~\cite{kieseberg2023controllable}. In recommendation systems, it allows users to refine item selections or add constraints, though excessive freedom may increase cognitive load and obscure how prompts map to outputs~\cite{bostandjiev2012tasteweights,di2018study,Satyanarayan2024Intelligence,wang2024promptcharm,jannach2017user}. Approaches include static preference profiles or dynamic updates via sliders or real-time visualizations~\cite{jannach2017user,harper2015putting}. For conversational AI, chaining prompts and requesting explanations help users shape responses and boost performance~\cite{wu2022ai,wei2022chain}. In educational AI, controllability remains underexplored despite its recognized importance \cite{aslan2024immersive,guan2023dilemma}. Aslan et al. used pre-defined templates to manage multi-modal pedagogical agents, but the impact of granting learners more control is unclear \cite{aslan2024immersive}. Therefore, this study investigates how controllability supports self-directed learning's planning phase.

%% file: sections/3-user_needs.tex
\section{User Needs and Design Requirements}
This section presents the findings from a formative study to identify user needs for LLM-based self-directed learning planning, leading to four key design requirements.

\subsection{User Needs}
To understand user needs, we conducted a formative study with 28 participants, each completing a 15-minute survey for a \$2 Amazon gift card. Ten of them also participated in follow-up interviews lasting 45 minutes, each receiving an \$8 Amazon gift card. Additionally, we interviewed an educational researcher. The study was approved by the Institutional Review Board at Texas A\&M University.

\subsubsection{Survey} 
The survey included 28 participants (19 females, 9 males), aged 19--29 years ($M = 25.37, SD = 2.38$), recruited through Slack and WeChat. The survey has 25 questions including multiple-choice, yes/no, open-ended, and 5-point Likert scale questions. It explored participants' experiences with self-directed learning, AI tools, and study plan generation. It also examined the frequency and effectiveness of AI tools like ChatGPT, control over AI-generated results, explanations, and desired improvements in study planning. Two researchers analyzed open-ended responses using affinity diagramming~\cite{hartson2012ux}, while quantitative analysis focused on descriptive statistics and response frequencies.

The results showed that of 26 participants (92.86\%) who reported trying AI tools such as ChatGPT or Notion AI, 10 (38.46\%) specifically used them for generating study plans. Among these, 5 participants (50\%) used the tools occasionally, while 3 participants (30\%) used them exclusively before exams. However, 9 out of 10 participants (90\%) encountered issues with AI-generated study plans, reporting errors or inaccuracies such as presenting incorrect information, misunderstanding questions, or providing vague explanations without evidence. Participants also noted inconsistencies in study plans such as mismatched resources that are not aligned or instances where the AI misinterpreted context or abbreviations with multiple meanings which confused them. As a result, participants felt that AI-generated plans often require human intervention, with 40\% stating corrections were needed frequently, 30\% always, and 30\% occasionally. While 8 participants (80\%) found explanations provided in their previous experiences with AI somewhat helpful, 9 participants (90\%) expressed a desire for more detailed reasoning behind AI recommendations. Additionally, participants expressed mixed-to-negative opinions about the ease of using the tools, with 40\% finding them not easy, 30\% feeling neutral, and 30\% finding them somewhat easy. They emphasized the need for greater control over modifying plans and requested features such as alternative study resources, enhanced interfaces, and iterative editing capabilities to better personalize their learning experience.

\subsubsection{Semi-structured interviews}
We interviewed 10 participants (9 female, 1 male), aged 22--29 years ($M = 25.7, SD = 1.95$). Participants were asked to use GPT-4o to create personalized study plans by writing their own prompts (e.g., ``\textit{I want to study GraphQL within 2 weeks using books, YouTube, and technical blogs with the goal of deploying a website.}'') and could modify and re-prompt as needed. After that, they provided feedback on the generated plans and shared views on GPT-4o's potential for personalized study systems, AI explainability, user control, and suggested improvements. Interview questions are listed in Table~\ref{SemiInterviewQuestions}.

The authors employed a systematic qualitative analysis following established practices~\cite{miles2014qualitative}. Interview responses were categorized into feedback on the current system and suggestions for future systems. Open-coding~\cite{charmaz2006constructing} and affinity diagramming~\cite{hartson2012ux} were used for analysis. Two researchers collaboratively developed a codebook\footnote{\url{https://osf.io/mfv8j?view_only=85c27a46d83e4d78864e26c2e0bf1fa5}} using responses from three randomly selected participants and independently analyzed the rest, resolving discrepancies through discussion. The current system feedback highlighted positive aspects, such as comprehensiveness, clarity, and personalization, alongside limitations like resource accessibility issues, hallucinations, inconsistent topic connections, low trust, limited control, insufficient personalization, and inadequate explanations. Users emphasized the need for better explanations of topic relevance and purpose. 
Future system improvements centered on controllability, reliability, materials, and information reception. For controllability, users desired features like iterative editing and flexible time adjustments. Reliability concerns included resource validation and clear reasoning, with explanations for topic sequencing. Participants sought diverse, accessible resources, including alternative videos for varied learning styles. For information presentation, users preferred customizable formats such as toggle-able sections for detailed explanations.
The analysis achieved a Cohen's Kappa of 0.89, indicating strong inter-rater reliability.

\begin{table}[h!]
    \centering
    \small
    \begin{tabular}{p{0.9\linewidth}}
        \toprule
        \textbf{User Satisfaction}\\
        Q1: How satisfied are you with the results? \\
        Q2: Where did you find challenging?\\
        Q3: Which parts of the recommendations are most helpful? \\
        Q4: Which parts do you find less helpful? \\
        Q5: Do the suggested study plans and materials align with your expectations and needs? \\
        Q6: Do you think LLM’s suggestions make sense to you? \\
        Q7: Do you trust the results? \\
        Q8: Do you want to take this plan? \\
        \midrule
        \textbf{User Experience: Controllability} \\
        Q9: Do you feel you have enough control? \\
        Q10: How was your experience re-prompting to refine the plan? \\
        \textbf{User Experience: Explainability} \\
        Q11: Do you need explanations for suggestions in study plans? \\
        Q12: Did you get enough explanations? \\
        Q13: What other explanations did you want to get? \\
        Q14: How clear were the reasons behind the suggestions? \\
        Q15: How do you prefer explanations to be provided (e.g., concise, detailed), and in what format (e.g., icons)? \\
        \midrule
        \textbf{Future Development} \\
        Q16: Would explanations help you prompt better? \\ 
        Q17: What types of study materials do you prefer to use for learning? \\
        Q18: Do you want to have alternative study materials? How many of them would you like? \\ 
        Q19: What criteria do you use for alternative study materials? \\
        Q20: Do you have any suggestions for future development? \\
        \bottomrule
    \end{tabular}
 \caption{Interviews focus on three aspects: user satisfaction (Q1-Q8), user experience (controllability: Q9-Q10, explainability: Q11-Q15), future development (Q16-Q20).}
        \label{SemiInterviewQuestions}
\end{table}

\subsubsection{Interview with an educational researcher}
We also conducted a one-hour Zoom interview with an educational researcher at the Center for Teaching Excellence at Texas A\&M University specializing in curriculum and instruction. The goal was to gather expert insights on the system's alignment with established learning theories and evaluate the educational validity of user-requested features for future development. After presenting the GPT-4o generated study plans and discussing identified issues and user-requested features, the researcher emphasized the importance of integrating established learning theories into the study plan generation process such as Bloom's Taxonomy~\cite{uark_blooms_taxonomy} for clear learning objectives. The researcher also suggested specific learning theories for each stage of the process, which we detail in Section~\ref{sec:planglow_design}. Additionally, she supported enhanced controllability and explainability would strengthen users' ability to effectively self-direct their learning plans. 

\subsection{Design Requirements}
From surveys and interviews with participants (P1–P10) and an educational researcher (E1), we found that explainability and controllability were top priorities. Additional design requirements also emerged to address other needs identified in the formative study.

\paragraph{\textbf{D1: Providing Contextual Explanations of the Recommendations}}
Participants emphasized the need for clear explanations of the system's recommendations, including the rationale behind resource selection, video prioritization, and topic significance. Understanding the importance of each topic would help them better evaluate and follow the plan. P2, P3, P4, and P5 noted that as beginners in the subject, they often needed more detailed explanations, such as guidance on where to start. E1 supported these views, highlighting the necessity of clear explanations in educational reasoning. Previous studies also show that clear explanations enhance usability and help users understand and trust system recommendations~\cite{abu2024supporting,ooge2022explaining}. Participants also preferred progressive disclosure, where detailed explanations can be revealed or hidden based on their needs. For example, P6 suggested toggle-able sections to prevent cognitive overload while allowing deeper exploration when needed~\cite{conati2021toward}. Therefore, contextual and progressively disclosed explanations would improve user experience and support informed decision-making for study plans.

\begin{figure*}
    \centering
    \includegraphics[width=1\linewidth]{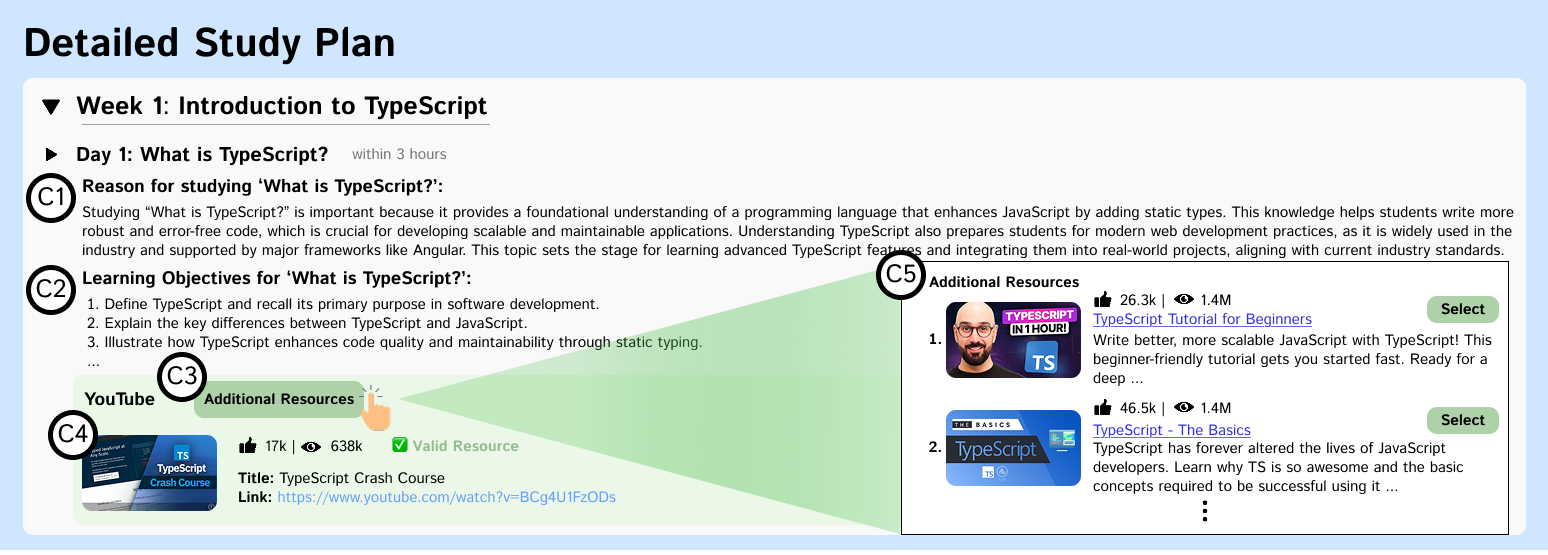}
    \caption{Detailed study plan of \textit{PlanGlow} organizes each week into five days. (C1) explains the reasons for studying each day's topic. (C2) lists learning objectives. (C3) allows users to explore additional resources via the button connecting to (C5). (C4) displays video resources with their status. A green check icon marks `Valid Resource', while a red icon indicates `Invalid Resource'. (C5) displays 10 additional resources with views, likes, and video descriptions. Clicking the `Select' button replaces the original plan's video with the selected one.}
    \label{fig:DetailedPlan}
\end{figure*}

\paragraph{\textbf{D2: Providing Reliable and Alternative Study Materials}} 
Participants valued accessible and reliable study materials to be recommended. They wanted materials with verifiable links, brief descriptions, and relevant metrics (e.g., likes, views) to ensure reliability and match their expectations. Mohammadi et al. also demonstrated that reliable resources increase user trust and engagement in learning~\cite{Mohammadi2019Trust}. Additionally, participants preferred alternative resources to accommodate different learning preferences and paces. P6 and P10 emphasized the need for comparative information between alternatives to make informed decisions on learning resources. E1 highlighted that diverse, reliable resources enhance user confidence, support learning styles, and foster trust for personalized plans.

\paragraph{\textbf{D3: Enhancing Controllability by Allowing Easy Plan Creation and Modification}} 
Participants prioritized the need for controllability in generating and customizing study plans to meet their specific needs. P2, P3, P4, P5, P6, and P10 expressed frustration with current systems that failed to provide satisfactory initial plans or made adjustments difficult. Research shows that study plan quality depends heavily on the input~\cite{lin2024exploring}. Incorporating structured input fields can guide users to provide precise responses, improving plan alignment. Participants also stressed the importance of easy, iterative modifications. Interactive features like clickable icons and editable fields empower users to refine plans in real-time, aligning them with learning goals. This aligns with Andragogy principles, which emphasize active involvement for adult learners. Empowering self-directed adults to customize plans fosters ownership, engagement, focus, and improved learning outcomes~\cite{knowles1978andragogy}.

\paragraph{\textbf{D4: Designing a User-Friendly and Clear Format for Study Plans}}
Clear formats and visuals improve comprehension and reduce cognitive load~\cite{ohlund2020valuable}. Participants (P1--P6, P8) emphasized breaking study plans into manageable sections, with lists to outline key topics and resources to enhance understanding and engagement. P5 suggested daily segments over weekly overviews for clarity, while P10 recommended interactive elements such as icons for easy customization. These features enhance usability and offer a more engaging and personalized experience.

%% file: sections/4-system.tex
\section{\textit{PlanGlow} System Design}\label{sec:planglow_design}

\textit{PlanGlow} is a web application consisting of two main components: the plan overview (Figure~\ref{fig:teaserl} (B1, B2)) and the detailed study plan (Figure~\ref{fig:DetailedPlan}). Its controllability features include \textbf{video replacement (D1, D2, D3)}, \textbf{input forms (D3, D4)}, \textbf{in-line editing (D3, D4)}, and a \textbf{chat feature (D3, D4)}, while its explainability features provide \textbf{background knowledge level descriptions (D1)}, \textbf{hierarchical explanations (D1, D4)}, and \textbf{video verification (D2)}. The application's front-end is developed using React.js with CSS styling, and the back-end is built with Python FastAPI. The back-end utilizes the OpenAI GPT-4o API to generate personalized study plans with explanations, such as content summaries, learning objectives, and reasons for resource selection. Detailed prompts can be found in the supplementary materials. The YouTube Data API v3 is integrated to recommend relevant and reliable resources.

To create personalized study plans, users start by filling out an input form in Figure~\ref{fig:teaserl} (A) with details such as the subject they want to study, learning goals, background knowledge, duration, and daily availability. This form enhances controllability by allowing users to customize their plans based on their needs and preferences. Its intuitive design improves usability by simplifying the process of providing necessary information, reducing cognitive load, and making the system accessible to a wide range of users, including those with no prior experience using prompt-based systems.

For users who are unsure about their current knowledge level, they can click the "bulb" \faIcon{lightbulb} icon in Figure~\ref{fig:teaserl} (A1) to see concise, informative descriptions of each level. This feature enhances usability by helping them confidently select the most appropriate level, ensuring the study plan is tailored to their abilities. The system leverages the OpenAI API to generate background knowledge insights based on the framework ~\cite{benner1982novice}, which categorizes expertise into six levels: novice, advanced beginner, competence, proficiency, expertise, and mastery. Each level is clearly explained in the prompt, defined by specific characteristics of learning and performance progression as outlined in the framework. The prompt is fine-tuned with parameters: \textit{temperature} (0.2) for factual accuracy, \textit{top\_p} (0.6) for likely completions, \textit{frequency\_penalty} (0.2) to reduce repetition, and \textit{presence\_penalty} (0.1) to encourage relevant new information.

\begin{figure*}
    \centering
    \includegraphics[width=1\linewidth]{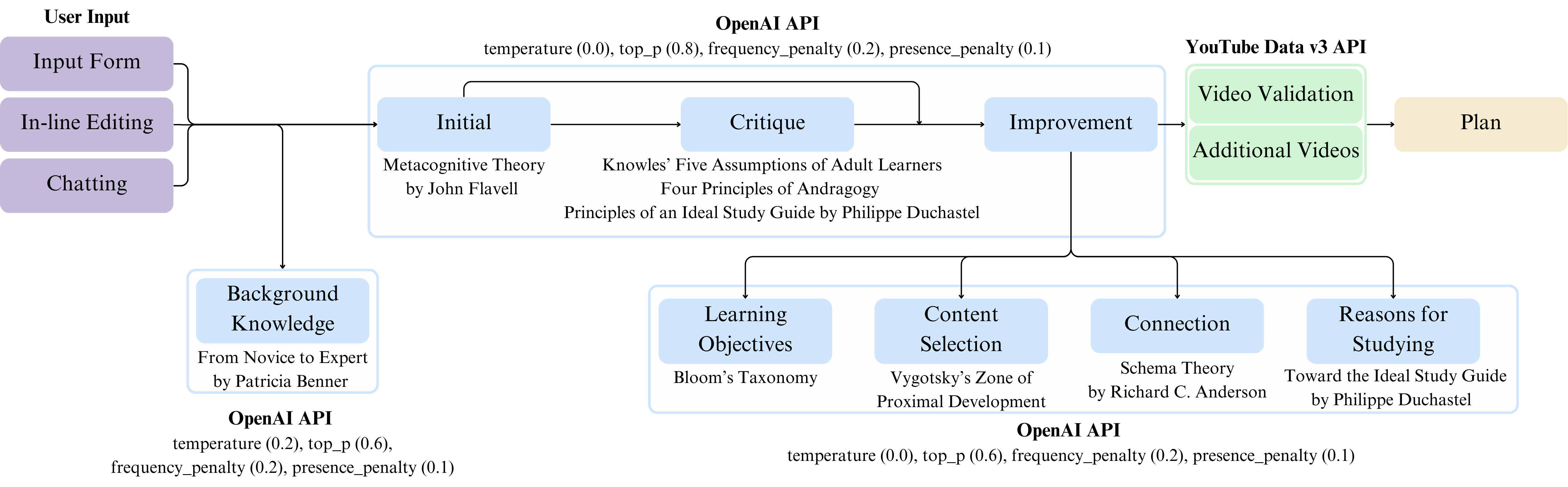}
    \caption{The workflow of \textit{PlanGlow} progresses from left to right. The system begins by collecting user inputs through an input form, in-line editing, or chat interface to create an initial plan and describe the background knowledge level. The study plan is generated through three sequential steps using the OpenAI API: Initial Generation, Critique, and Improvement. The final plan incorporates comprehensive elements, including learning objectives, content selection rationales, conceptual connections across daily and weekly units, and explanations for studying each topic. All video resources are validated and supplemented by the YouTube Data v3 API. Each block details the specific learning theories applied and parameters tuned.}
    \label{fig:system}
\end{figure*}

Once the form is submitted, the system generates a personalized study plan (Figure~\ref{fig:teaserl} (B)) using a three-step chain-of-thought ~\cite{wei2022chain} process with the OpenAI API, as outlined in Figure~\ref{fig:system}: the initial step, critique step, and improvement step. In the initial step, the system creates a draft study plan based on the user's form inputs. This step provides a coherent, week-by-week structure, integrating metacognitive strategies such as planning, monitoring, and evaluating learning. With these strategies, the system ensures the plan is practical, achievable, and support self-directed learning. This approach, grounded in metacognitive theory~\cite{brown1978knowing, duell1986mcskills, flavell1976metacognitive}, provides a theoretical basis for creating study plans that enhance learning outcomes. To further promote the user's engagement and skill development, the system breaks the plan into daily tasks that increase in complexity over time and recommends accessible and relevant resources. Together, these elements generate a cohesive initial draft, which is refined in the critique and improvement steps. 
Next, in the critique step, the system evaluates the initial study plan using adult learning theories ~\cite{knowles2014adult, knowles1977adult} and study guide principles ~\cite{duchastel1983toward}. This step applies Knowles' Five Assumptions of Adult Learners and the Four Principles of Andragogy to ensure the study plan supports self-directed learning, leveraging prior experiences, emphasizing practical approaches, and encouraging internal motivation. The critique also checks the plan's clarity, relevance, and organization based on study guide principles. Finally, in the improvement step, the system refines the initial study plan using feedback from the critique. Throughout the process of generating the study plan, the system uses tuned parameters to ensure accuracy and quality: \textit{temperature} (0.0), \textit{top\_p} (0.8), \textit{frequency\_penalty} (0.2), and \textit{presence\_penalty} (0.1). Users can further interact with \textit{PlanGlow} through in-line editing (Figure~\ref{fig:teaserl} (B1)) to easily adjust key elements such as background knowledge level, study subjects, goals, duration, or daily availability. Additionally, they can utilize the chat feature to edit the plan or to ask any questions. Plan edits follow the three-step chain-of-thought process in Figure~\ref{fig:system}, while other queries receive targeted responses tailored to the user's needs.

The generated plan follows a layered hierarchical approach, starting with a high-level overview of the study plan.  Users can expand weekly sections to view the learning objectives, reasoning behind the selected content, and the connections between daily materials in Figure~\ref{fig:teaserl} (B2). Further expansion reveals more detailed daily plans, balancing a broad understanding with in-depth exploration. This structure enhances usability by presenting complex information in manageable layers, from the overall plan to the finer details. The system employs Bloom's Taxonomy~\cite{uark_blooms_taxonomy} to create clear and structured learning objectives. Through the OpenAI API, each objective from foundational skills such as remembering and understanding, to advanced skills such as analyzing, evaluating, and creating, ensures a well-rounded and progressive learning experience. The system generates reasoning for resource selection based on Vygotsky's Zone of Proximal Development (ZPD) ~\cite{chaiklin2003zone} to match the learner's abilities and gradually increase complexity. It ensures diversity, accessibility, interactivity, and quality to support learners' growth and align with their goals. To enhance comprehension, the system connects relevant prior knowledge structures when introducing new content. This approach, grounded in schema theory~\cite{anderson2018role}, shows how learners use existing knowledge to interpret and integrate new information, enhancing understanding and retention.

For daily study plans, the system outlines the rationale for selecting topics (Figure~\ref{fig:DetailedPlan} (C1)), specific learning objectives (Figure~\ref{fig:DetailedPlan} (C2)). The prompt guides the system to explain the reasons for studying each topic by addressing learners' needs for clarity, orientation, and goal-setting. It highlights each topic's relevance within the study plan, its connection to broader course objectives, and its role in achieving learning goals~\cite{duchastel1983toward}. The system also presents reliable and relevant video resources with likes, views, and validation (Figure~\ref{fig:DetailedPlan} (C4)). It validates videos using the YouTube Data v3 API, includes valid videos in the study plan, and replaces invalid videos automatically with suitable alternatives matched to learner's topic, proficiency, and daily study time, sorted by ratings. To enhance usability, the system provides a clear visual indicator provides real-time feedback on video validity. In Figure~\ref{fig:DetailedPlan} (C4), a green check icon indicates valid resources labeled as `Valid resources,' while a red icon signifies invalid resources, ensuring reliability and enabling users to make adjustments as needed. In addition to the primary videos, users can click on Figure~\ref{fig:DetailedPlan} (C3) to view up to 10 additional video suggestions (Figure~\ref{fig:DetailedPlan} (C5)), ranked by relevance, with details such as views, likes, and descriptions. Users can select their preferred video to replace the original one in the study plan, offering flexibility to customize their learning experience. This enables users to tailor the plan to their individual preferences and goals, ensuring it effectively supports their learning process.

%% file: sections/5-evaluation.tex
\section{Evaluation of \textit{PlanGlow} System}
This section presents a within-subject user study evaluating \textit{PlanGlow}'s performance, usability, and user experience, with a focus on controllability and explainability. The protocol was approved by the Texas A\&M University's Institutional Review Board.

\subsection{Participants}
We recruited 24 current students from Texas A\&M University (9 males, 15 females; 15 Bachelor's, 7 Master's, 2 Ph.D.), all of whom had experience with self-directed learning. A power analysis confirmed the sample size was sufficient for a large effect (\( d = 0.8 \), \( \alpha_{\text{Bonferroni}} = 0.05 / 3 \), power = 0.9). Of the participants, 21 were familiar with LLMs like ChatGPT, and 10 had used AI for study planning. The study, conducted via Zoom, lasted 1--1.5 hours per session, with participants accessing the systems as web applications. Each received a \$20 Amazon gift card as compensation.

\begin{figure}
    \centering
    \includegraphics[width=1\linewidth]{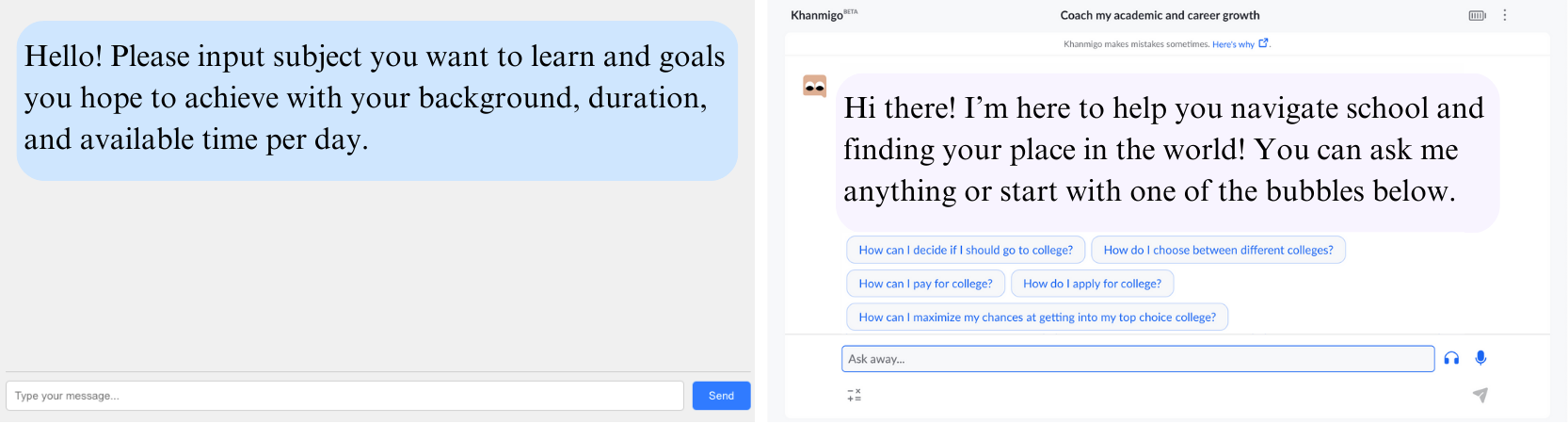}
    \caption{The interface of two systems are compared with \textit{PlanGlow} in the evaluation: a GPT-4o-based system (left) and Khan Academy's Khanmigo for the `Coach my academic and career growth' feature (right).}
    \label{fig:Khan}
\end{figure}

\begin{table}[h!]
\centering
\small
\begin{tabular}{p{0.9\linewidth}}
\toprule
\textbf{Performance} \\
Q1: The study plan was clearly presented. \\
Q2: The study plan met my goals and needs. \\
Q3: I want to use this study plan in my learning process. \\
\midrule
\textbf{Usability} \\
Q4: I would like to use this system frequently. \\
Q5: The system is unnecessarily complex. \\
Q6: The system was easy to use. \\
Q7: The various functions in the system were well integrated. \\
Q8: There was too much inconsistency in the system. \\
Q9: I would imagine that most people would learn to use this system very quickly. \\
Q10: I felt very confident using the system. \\
\midrule
\textbf{User Experience: Controllability} \\
Q11: The system allows me to generate the study plan easily. \\
Q12: The system efficiently provides the desired study plan.\\
Q13: The process of searching for additional or alternative study resources is straightforward. \\
Q14: The system ensures that the process of verifying the validity of study resources is reliable. \\
Q15: I feel confident and able to revise the study plan easily to suit my specific goals and needs. \\
\midrule
\textbf{User Experience: Explainability} \\
Q16: The system provides concise explanations. \\
Q17: The explanations offered by the system are accurate. \\
Q18: The explanations provided by the system are relevant. \\
Q19: The system helps me understand why certain recommendations or resources were included. \\
Q20: The system explains how the study plan aligns with my goals and preferences. \\
Q21: The system clearly explains the connection between daily and weekly study tasks. \\
Q22: The system offers explanations that help me make informed decisions about following or revising the study plan. \\
\bottomrule
\end{tabular}
\caption{Survey focuses on three dimensions: performance (Q1-Q3), usability (Q4-Q10), user experience (controllability: Q11-Q15, explainability: Q16-Q22).}
\label{SurveyQuestions}
\end{table}  
\subsection{Baselines}
The study compares \textit{PlanGlow} to two baseline systems: a GPT-4o-based system~\cite{fan2024lessonplanner} and Khan Academy's Khanmigo\footnote{\href{https://www.khanmigo.ai/}{https://www.khanmigo.ai/}}. GPT-4o is the commonly used system for generating personalized feedback and has been used as the baseline for generating curriculum for teachers~\cite{fan2024lessonplanner}. Khanmigo is a popular commercial tool for generating study plans. The GPT-4o-based system has a simple interface where users are asked to "\textit{Please input the subject you want to learn and goals you hope to achieve with your background, duration, and available time per day.}" in a text box (Figure~\ref{fig:Khan}). For Khanmigo, we use the `Coach my academic and career growth' feature in Figure~\ref{fig:Khan}. To ensure balanced testing, six possible system usage orders were assigned, with four participants per order.

\subsection{Tasks} Before using each system, participants were introduced to its main features. They were asked to create personalized study plans on subjects of their choice using each system, without field restrictions. Participants iteratively create and refine their study plans as needed. Once they were satisfied with the generated plan from each system, they completed a post-task survey, explaining the reasons for their responses to each question. The survey included 22 identical questions, all based on a 7-point Likert scale (1 = Strongly Disagree, 7 = Strongly Agree) assessing performance, usability, and user experience, with a focus on controllability and explainability. A full list of the survey questions is provided in Table~\ref{SurveyQuestions}. Usability questions were adapted from the System Usability Scale and user experience questions were based on the previous framework \cite{wang2024promptcharm}. After using all three systems, participants were asked which system they preferred the most.

\subsection{Hypothesis} We propose the following hypotheses based on prior work ~\cite{xia2019peerlens}. \\
\indent{\textbf{H1.}} \textit{PlanGlow} outperforms baseline systems in performance by improving study plan clarity (H1a), alignment with user goals and needs (H1b), and user intention to adopt the plan (H1c).
\\ \indent{\textbf{H2.}} \textit{PlanGlow} demonstrates better usability over baseline systems by encouraging frequent use (H2a), reducing complexity (H2b), enhancing ease of use (H2c), integrating functions effectively (H2d), ensuring consistency (H2e), supporting quick learnability (H2f), and boosting user confidence (H2g).
\\ \indent{\textbf{H3.}} \textit{PlanGlow} offers greater controllability by enabling effortless study plan generation (H3a), efficiently achieving desired plans (H3b), effective resource search (H3c), reliable resource validation (H3d), and confident plan revisions to meet goals and needs (H3e).
\\ \indent{\textbf{H4.}} \textit{PlanGlow} provides more effective explanations, offering concise (H4a), accurate (H4b), and relevant (H4c) information. It helps users understand the rationale behind recommendations (H4d), aligns plans with user goals (H4e), clarifies connections between daily and weekly study tasks (H4f), and empowers informed decisions on following or revising plans (H4g).

\subsection{Evaluation of Study Plans}
We recruited two educators: E1, a female assistant professor in special education specializing in learning development at the Nanjing Normal University of Special Education, and E2, a male educational consultant at the Texas A\&M University Center for Teaching Excellence. 
E1 and E2 collaboratively developed evaluation criteria for five standardized questions that incorporate a 5-point Likert scale (1 = Strongly Disagree, 5 = Strongly Agree), adapted from prior works~\cite{li2010factors, du2012using}. The full questions are in Table~\ref{tab:evalMatrix} and the evaluation criteria are available online\footnote{\url{https://osf.io/bwjqg?view_only=85c27a46d83e4d78864e26c2e0bf1fa5}}.
Based on these criteria, E1 evaluated all generated plans from the user study. The plans were presented as text in a counterbalanced order along with each participant's context (subject, background knowledge, study duration, daily availability), while \textit{PlanGlow} results were simplified to include only the Week 1 overview  (Figure~\ref{fig:teaserl} (B1, B2)) and Day 1 detailed plan (Figure~\ref{fig:DetailedPlan}). E1 was informed that full explanations covered all weeks and days, and that resource availability had been confirmed but was not shown in the presented plans. Further details on how the plans were presented are available online\footnote{\url{https://osf.io/q3uf4?view_only=85c27a46d83e4d78864e26c2e0bf1fa5}}.

\subsubsection{Hypothesis} 
We hypothesize \textbf{H5} that \textit{PlanGlow} enhances educational quality with clearer learning objectives (H5a), more accurate timelines (H5b), better-supported activities with reliable resources (H5c), more effective progress monitoring (H5d), and stronger pedagogical foundations (H5e) compared to the baselines.

\begin{table}[h!]
\centering
\small
\begin{tabular}{p{\linewidth}}
    \toprule
    Q1: Does the plan outline clear and specific learning objectives? \\
    Q2: Does the plan include a timeline or estimated completion time? \\
    Q3: Does the plan detail practical, well-supported learning activities that are feasible and make use of available resources? \\
    Q4: Does the plan provide methods to monitor and measure progress toward achieving the learning goals? \\
    Q5: Is the plan and its explanation pedagogically sound? \\
    \bottomrule
\end{tabular}
\caption{Evaluation matrix for assessing generated plans.}
\label{tab:evalMatrix}
\end{table}

%% file: sections/6-results.tex
\section{Results} 
\begin{figure}
    \centering
    \includegraphics[width=1\linewidth]{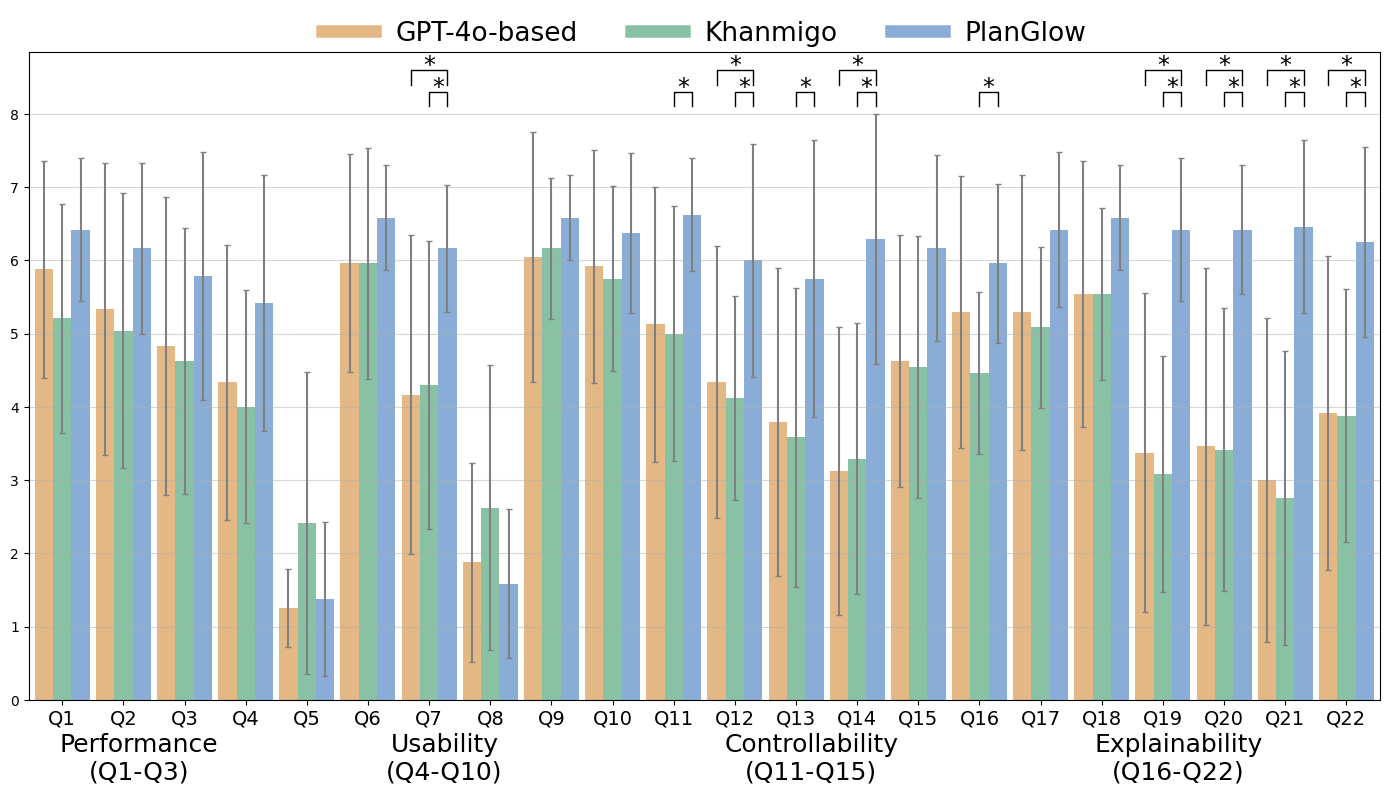}
    \caption{Means and standard errors of GPT-4o-based system, Khanmigo, and \textit{PlanGlow} on a 7-point Likert scale ($\ast$: $p < .05$).}
    \label{fig:UserStudyResult}
\end{figure}

\begin{figure}
    \centering
    \includegraphics[width=1\linewidth]{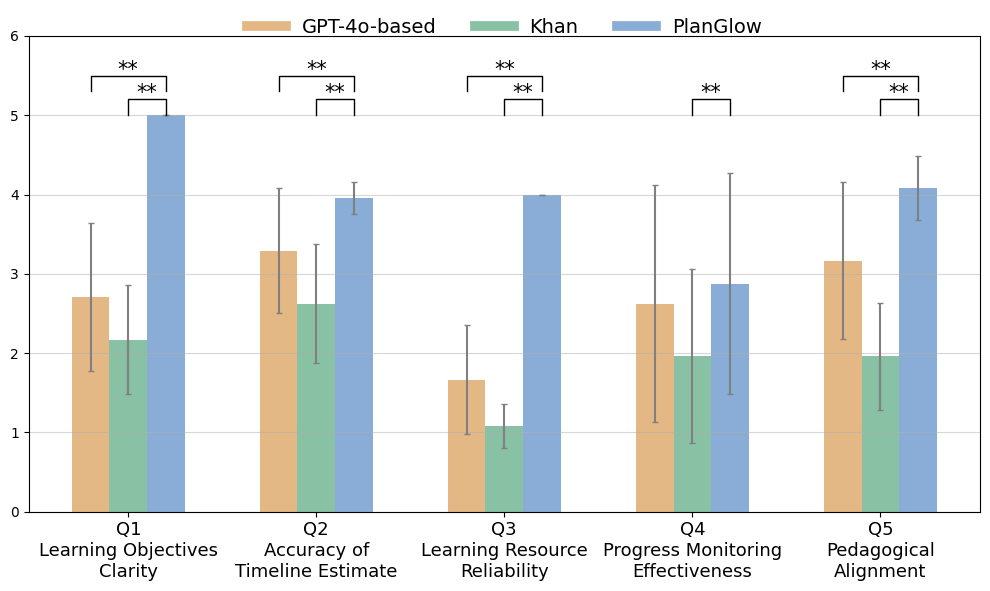}
    \caption{Means and standard errors for each evaluation assess the educational quality of generated plans on a 5-point Likert scale ($\ast\ast$: $p < .01$).}
    \label{fig:EduEval}
\end{figure}

This section presents findings from participant feedback from the user study (P1--P24) and educational evaluations of \textit{PlanGlow} and two baseline systems. Figure~\ref{fig:UserStudyResult} details survey results, with statistical significance assessed using ANOVA with Bonferroni post-hoc tests. Figure~\ref{fig:EduEval} shows experts' evaluations of generated plans, analyzed using One-way ANOVA.

\paragraph{\textbf{Participant Behavior and Learning Preferences}}
Participants' learning subjects fell into six categories: 29.2\% focused on language learning (e.g., Spanish, Japanese), 20.8\% on programming or technology (e.g., Typescript, Python), 20.8\% on academic topics (e.g., UAVs, WJ IV intellect assessment), 12.5\% on research and writing, and 8.3\% on creative skills (e.g., video editing, dancing). Background knowledge levels included 29.2\% novices, 41.7\% advanced beginners, 20.8\% competent, 4.2\% experts, and 4.2\% mastery. Interaction data showed all participants viewed the background knowledge description once, submitted plans 1.67 times, and used chat features 0.63 times on average. Two participants used in-line editing, making two modifications each. They accessed additional resources 6.13 times, selecting 3.5 on average, and viewed weekly and daily explanations 13.042 and 10.33 times, respectively.

\paragraph{\textbf{Performance (H1)}} 
We do not notice significant performance differences across the systems. \textit{PlanGlow} had higher means and lower standard deviations for study plan clarity ($M = 6.42, SD = 0.97$), alignment with user goals and needs ($M = 6.17, SD = 1.17$), and adoption ($M = 5.79, SD = 1.69$) compared to GPT-4o and Khanmigo, but results were not statistically significant. Thus, \textbf{H1a-H1c are rejected}. Participants described \textit{PlanGlow}'s comprehensive and well-organized plans, highlighting clear goals and progressive learning paths (P9, P13, P15, P19, P24). P10 also noted alignment with a previous class she had taken. However, limitations such as slow response time (P6, P9, P16) and broad resources (e.g., full tutorials) reduced its advantage over baselines.

\paragraph{\textbf{Usability (H2)}}
Most usability hypotheses showed no significant differences between \textit{PlanGlow} and the baselines ($p > .05$, \textbf{H2a-H2c, H2e-H2g rejected}). However, \textit{PlanGlow} significantly outperformed both GPT-4o ($p < .01$) and Khanmigo ($p < .05$) in functional integration, \textbf{supporting H2d}. This improvement is attributed to \textit{PlanGlow}'s comprehensive interface, offering multiple interaction methods beyond the chat-based focus of the baselines. The rejections of other hypotheses may reflect similar underlying LLMs across systems and the short-term study duration, limiting the exploration of nuanced usability differences. 

\paragraph{\textbf{User Experience: Controllability (H3)}}
Participants reported greater controllability with \textit{PlanGlow} compared to the baselines. \textit{PlanGlow} facilitated easier study plan generation than Khanmigo($p<.05$, \textbf{H3a supported}) and was significantly more efficient in delivering desired plans than both GPT-4o and Khanmigo ($p<.05$, \textbf{H3b supported}). It also simplified searching for additional resources compared to Khanmigo ($p<.05$, \textbf{H3c supported}) and showed greater reliability in resource validation than both GPT-4o and Khanmigo ($p < .01$, \textbf{H3d supported}). However, no significant difference was found in capabilities to revise study plans to suit their specific goals and needs (\textbf{H3e rejected}). Participants mentioned \textit{PlanGlow}'s structured input form for easier plan generation (P1, P9, P16, P19, P22) and its flexibility in customizing plans especially replacing video resources to fit their learning preferences and goals, P10 noted, ``\textit{The ability to search and replace study resources is very helpful. Previously, when studying this topic, I had a favorite YouTuber, and now I am glad that I can include their video that fits perfectly here.}'' In contrast, Khanmigo's chat-based input was noted as challenging (P5, P10, P23). Despite these strengths, similar basic editing features across systems likely explain the lack of significant differences in plan revision.

\paragraph{\textbf{User Experience: Explainability (H4)}}
\textit{PlanGlow} showed a significant advantage in providing concise explanations compared to Khanmigo ($p<.05$, \textbf{H4a supported}) but no significant differences in accuracy or relevance (\textbf{H4b, H4c rejected}). \textit{PlanGlow} outperformed both GPT-4o and Khanmigo in explaining the rationale behind recommendations ($p < .01$, \textbf{H4d supported}), aligning study plans with user goals ($p < .01$, \textbf{H4e supported}), and clarifying connections between daily and weekly tasks ($p < .01$, \textbf{H4f supported}). \textit{PlanGlow} also empowered users to make informed decisions about study plans compared to both baselines ($p < .01$, \textbf{H4g supported}). P24 stated, ``\textit{The explanations are concise, easy to read, understand, and intuitive.}'' P3 added that, ``\textit{I like the concise explanations in the overview.}'' Similarly, P1 commented, ``\textit{I like learning objectives in the overview and detailed plan because they help me understand what I need to know.}'' P11 highlighted, ``\textit{I can easily understand why these are recommended to me, and they align well with my intentions.}'' While all participants reviewed and approved the accuracy and relevance of the explanations, no significant differences were observed. Supporting this, P10 trusted Khanmigo's explanations due to its reputation as a reliable educational website.

\paragraph{\textbf{Overall Preferences}} 
A total of 83.3\% of participants (20) ranked \textit{PlanGlow} as their top choice, compared to 12.5\% (3) for Khanmigo and 20.8\% (5) for GPT-4o. Three participants provided tied rankings: P8 and P20 preferred both \textit{PlanGlow} and GPT-4o, while P10 ranked all three systems equally.

\paragraph{\textbf{Study Plan Quality (H5)}}

\textit{PlanGlow} significantly outperformed the baselines. Compared to Khanmigo, \textit{PlanGlow} excelled in all aspects ($p < .01$, \textbf{H5a-H5e supported}). 
When compared to GPT-4o, it showed improvements in learning objectives, 
timeline estimates, reliable resources, and pedagogical principles
($p < .01$, \textbf{H5a-H5c, H5e supported}), while progress monitoring showed no significant difference ($p > .05$, \textbf{H5d rejected}).

%% file: sections/7-discussion.tex
\section{Discussion}
In this section, we discuss the design considerations, limitations, and future work.
\subsection{Design Considerations}
\paragraph{\textbf{DC 1: The Need for Enhanced Explainability}}
Our study demonstrates strong user demand for explainable AI in educational planning. The interaction logs reveal substantial engagements with explanation features. Participants viewed weekly explanations 13.04 times and daily explanations 10.33 times on average, accessed additional resources 6.13 times, and selected 3.5 resources per session. These high engagement rates with explanatory features align with prior work highlighting the importance of AI explainability in education~\cite{chaushi2023explainable, xu2020dilemma}. The frequent access to explanations suggests that users value understanding the AI's reasoning, particularly in plan structure and resource recommendations.

\paragraph{\textbf{DC 2: The Need for Enhanced Controllability}}
The study revealed important insights about user preferences regarding control in AI-assisted learning plan generation. Participants valued the structured input form (Figure~\ref{fig:teaserl} (A)) for straightforward plan generation, aligning with Satyanarayan et al. ~\cite{Satyanarayan2024Intelligence}, which suggests user preference for guided control over full autonomy in AI interactions. However, participants made limited use of additional control features. Specifically, only two participants utilized in-line editing, two edits each, and chat-based modifications averaging 0.63 per person. This low usage may reflect the one-time nature of the study, where participants may not have needed to significantly modify their plans. Future longitudinal studies could better evaluate the utility of plan modification features by examining how users adapt and update their learning plans over extended periods.

\paragraph{\textbf{DC 3: Accountability and Pedagogical Responsibility}}
The integration of AI in educational guidance raises important ethical considerations that \textit{PlanGlow} explicitly addresses. First, we maintain human agency by designing the system as a supportive tool rather than a replacement for educational expertise~\cite{porayska2023ethics, holmes2022ethics}. \textit{PlanGlow}'s explanations enable learners to make informed decisions about accepting or modifying recommendations which ensure accountability~\cite{luckin2016intelligence}. Second, we address pedagogical responsibility by incorporating established educational principles and maintaining transparency about its capabilities and limitations~\cite{holmes2022ethics}. \textit{PlanGlow} informs that its recommendations are suggestions rather than definitive instructions, encouraging critical evaluation by users. By allowing iterative refinement of plans, \textit{PlanGlow} supports a balanced approach where AI assists but does not dictate educational decisions, preserving the essential role of human judgment in learning. While our current evaluation has not identified specific ethical concerns, we acknowledge potential future challenges, particularly when dealing with sensitive subjects or specialized domains. As future work, we plan to investigate safeguards and guidelines for handling potentially sensitive educational content and ensure responsible AI-guided learning across diverse subject matters~\cite{akhtar2024socially}.

\subsection{Limitation and Future Work}
However, this work still has several limitations. \textit{PlanGlow} faces challenges in generalizing effectively across diverse domains due to the limitations of the underlying LLMs. These models are trained on vast datasets that may embed inherent biases or lack domain-specific knowledge, potentially leading to inaccurate plans in certain areas. 
A particular challenge emerges with specialized college-level subjects, where the system's performance is hindered by the scarcity of structured learning resources and detailed study plans in public datasets. For example, during user studies, highly domain-specific topics' plans such as UAVs or WJ IV intellect assessment lacked concrete implementation steps and sufficient domain expertise. While general educational content exists, resources for advanced specialized topics were limited.
To address these limitations, we will implement two main solutions. First, we will utilize specialized academic datasets from MIT OpenCourseWare~\cite{mit_opencourseware} and expert-validated study guides from Coursera~\cite{coursera_platform}. Second, we will develop domain-specific prompting strategies in collaboration with subject matter experts. These combined strategies aim to mitigate biases, improve scalability, and ensure \textit{PlanGlow} can provide accurate and adaptable plans across various domains.

Another limitation is that \textit{PlanGlow} relies solely on YouTube videos as the primary resource. This design choice reflects our focus on enhancing the system's controllability and explainability in its current iteration, rather than developing a comprehensive tool that integrates multiple resources. While this approach is effective for certain users, it limits the diversity of available resources, particularly for those who prefer other formats such as reading materials and exercises. Evaluation results from two educators highlighted this limitation by rejecting H5d against GPT-4o, which questioned whether \textit{PlanGlow} can monitor and measure progress toward achieving learning goals. Since \textit{PlanGlow} is designed primarily to assist with the planning phase of self-directed learning, it lacks features to support learners during the execution and monitoring of their progress. To address these issues, we aim to include diverse resources, such as assessments and interactive exercises, and to implement methods for monitoring and measuring learners' progress. These enhancements will involve strategies to track progress toward achieving learning goals and to strengthen learners' confidence in their self-directed learning abilities~\cite{li2010factors}. 

The last limitation was the skewed gender distribution during interviews, with most participants being female, potentially biased feedback. Future studies should ensure balanced gender representation for more inclusive insights into system design. 

%% file: sections/8-conclusion.tex
\section{Conclusion}
In this paper, we present \textit{PlanGlow}, a personalized study planning system offering explainable and controllable recommendations. Through a within-subject user study with 24 participants, we found that \textit{PlanGlow} significantly outperformed two baselines. In controllability, \textit{PlanGlow} demonstrated great performance in plan generation efficiency, resource management, and validation. For explainability, \textit{PlanGlow} significantly improved users' understanding of recommendation rationales, goal alignment, and task connections, enabling more informed decision-making. \textit{PlanGlow} also improved in study plan quality, including objectives, timelines, resources, monitoring, and pedagogy. While usability differences were minimal, \textit{PlanGlow} achieved notably better functional integration, with 83.3\% of participants ranking it as their preferred system. These findings highlight the transformative potential of integrating explainable and controllable AI in self-directed learning and education, offering a path toward more personalized, effective, and user-centered learning environments.